\documentclass[11pt,a4paper]{article}
\usepackage{algorithm}
\usepackage{algpseudocode}
\usepackage{amsmath, amssymb}
\usepackage[
	style=authoryear,
	maxbibnames=100,
]{biblatex}
\usepackage{datetime}
\usepackage{graphicx}
\usepackage{float}
\usepackage{listings}
\usepackage{mathtools}
\usepackage{placeins}
\usepackage{subcaption}
\usepackage{tikz}
\usepackage{url}

\usetikzlibrary{calc}

\newcommand{\pkg}[1]{\texttt{#1}}
\newcommand{\code}[1]{\texttt{#1}}
\newcommand{\proglang}[1]{\textsf{#1}}

\DeclareMathOperator*{\argmin}{arg\,min}

\title{Scentree: a framework for generating scenario trees for multistage stochastic programming}
\author{
	Cristian Pachón-García, Albert Solà Vilalta and F.-Javier Heredia\\
	{\small Departament d'Estadística i Investigació Operativa}\\
	{\small Universitat Politècnica de Catalunya}
}
\date{\monthname\ \the\year}

\begin{document}

\maketitle

\begin{abstract}
We present scentree, an open-source Python package for constructing scenario trees for multistage stochastic programming models using either historical data or samples drawn from a probability distribution. It combines machine learning and multivariate time series models to obtain a scenario fan that captures dependencies in the stochastic processes. This scenario fan is subsequently reduced into a scenario tree that preserves representativeness of the stochastic process, providing a flexible and extensible framework for uncertainty modeling in stochastic programming problems. The framework can also be naturally applied to two-stage stochastic programming problems. Key contributions are the joint modeling of the stochastic process, preserving correlations between all random variables, and the automation of the complete workflow, including forecasting model selection, parameter estimation, scenario fan generation, and scenario tree construction. In addition, scentree does not assume any underlying data distribution, reducing the statistical expertise required to produce a scenario tree, and is modular by design, allowing external forecasts and other partial executions of the workflow. 

\end{abstract}

\section{Introduction}
Uncertainty arises naturally in many situations where decision must be made before all relevant information is available. For instance, in finance, investment decisions are made under uncertainty about future market conditions \parencite{Dixit2012}. Similarly, in energy system operation, decisions regarding electricity generation must be made before future demand is known \parencite{Roald2023}. In these and other applications, uncertain quantities evolve over time, and new information becomes available as the decision-making process progresses.

Multistage stochastic programming \parencite{Birge2011,Pflug2014} provides a mathematical framework for addressing these decision problems by explicitly incorporating the evolution of uncertain parameters across multiple stages. At each stage, decisions can be adjusted according to the information available up to that point, allowing the model to account for the sequential revelation of uncertainty.

A key component of a multistage stochastic program is the representation of the underlying uncertainty. In its most general form, uncertain quantities can be described by a stochastic process. Usually, these stochastic processes admit infinitely many possible realizations, making their direct incorporation into a mathematical optimization model impractical. Consequently, a discrete approximation of the stochastic process is required. Such an approximation can be represented by a finite collection of scenarios \parencite{Dupacova2000}, where each scenario corresponds to one possible realization of the stochastic process over the planning horizon.

In multistage problems, scenarios must capture the structure of the underlying uncertain process, which is typically defined over an index set representing time or other relevant dimensions. A scenario fan consists of a collection of possible trajectories of the uncertain process, with each scenario representing one possible evolution across the stages. A scenario tree provides the structured representation of these scenarios following the multistage decision-taking process defined in the associated stochastic programming problem, where scenarios can share nodes at a given stage and branch into different possible subsequent evolutions. Probabilities are assigned to the nodes to quantify the likelihood of the corresponding realizations, allowing the resulting discrete representation to be incorporated into a multistage stochastic programming model.

The construction of representative scenarios and, subsequently, a suitable scenario tree is therefore an essential step in the practical application of multistage stochastic programming. A scenario representation should capture the main characteristics of the underlying stochastic process. When historical observations of the uncertain process are available, scenario generation can be approached directly from data rather than by assuming a predefined probability distribution. This approach is particularly relevant when the underlying process is multivariate or exhibits complex temporal dependencies.

Building on these considerations, we present \pkg{scentree}, an open-source \proglang{Python} \parencite{PythonFoundation} framework for generating scenario fans and scenario trees for multistage stochastic programming problems from historical data. The package is equally applicable to both two-stage and multistage stochastic problems. However, this work focuses on the multistage setting, as it is more general, complex, interesting and challenging from a methodological perspective, than two-stage problems, and fewer software tools are available for it.

In this context, the main contributions are (i) a data-driven approach based on historical data for developing predictive models; (ii) the integration of statistical and machine learning models; (iii) the ability to account for dependencies, such as correlations, among variables; (iv) a modular design supporting external forecasting tools; and (v) a fully automated workflow that reduces the expertise required from users to develop and apply forecasting algorithms.

To establish the mathematical framework used throughout the paper, we introduce the notation for the decision process, the optimization problem, the underlying stochastic process, and the scenario tree. Multistage stochastic programming models sequential decision-making under uncertainty, in which decisions are made as random information is revealed across stages. Following \textcite[][Section 1.2]{Pflug2014}, we represent this sequential structure as
\[
    \mathbf{x}_0 \longrightarrow \boldsymbol{\xi}_1 \longrightarrow \mathbf{x}_1 \longrightarrow \cdots \longrightarrow \mathbf{x}_{s-1} \longrightarrow \boldsymbol{\xi}_s,
\]
possibly followed by a terminal decision $\mathbf{x}_s$. The initial decision, $\mathbf{x}_0$, is deterministic, and each subsequent decision $\mathbf{x}_i$, $1 \leq i \leq s$, depends only on the observed random variables $(\boldsymbol{\xi}_1, \ldots, \boldsymbol{\xi}_i)$, ensuring nonanticipativity.

A multistage stochastic programming problem optimizes this decision process; in the risk-neutral case, its general formulation is
\begin{equation}
    \begin{aligned}
        \min_{\mathbf{x}\ \mathrm{nonanticipative}} \mathbb{E}\!\left[f(\mathbf{x},\boldsymbol{\xi})\right]
        \quad\text{subject to}\; \mathbf{x} \in \mathcal{X}(\boldsymbol{\xi})\quad\text{almost surely},
    \end{aligned}
\label{eq:general_msop}
\end{equation}
where $\mathbf{x}$ denotes the sequence of decisions, $\boldsymbol{\xi} \coloneqq(\boldsymbol{\xi}_1,\ldots,\boldsymbol{\xi}_s)$ the sequence of random variables, $f$ the total cost, and $\mathcal{X}(\boldsymbol{\xi})$ the feasible set. We model the uncertainty through a multivariate stochastic process
\[
    \boldsymbol{\xi} \colon \mathcal{D} \times \Omega \to \mathbb{R}^p,
\]
where $\mathcal{D}$ is the index set and $\Omega$ is the sample space. Each stage \(i\) is associated with a stochastic process
\[
    \boldsymbol{\xi}_i \colon \mathcal{D} \times \Omega \to \mathbb{R}^{p_i}
\]
with
\[
    \sum_{i = 1}^s p_i = p.
\]

Our objective is to obtain a discrete approximation $\boldsymbol{\xi}_{\mathrm{tr}}$ of the underlying stochastic process $\boldsymbol{\xi}$, represented by a scenario tree. We define
\[
    \boldsymbol{\xi}_{\mathrm{tr}} \colon \mathcal{D} \times \Omega_{\mathrm{tr}} \to \mathbb{R}^p,
\]
where $\Omega_{\mathrm{tr}}$ is a finite sample space whose elements correspond to the scenarios of the tree. A discrete probability distribution is defined over $\Omega_{\mathrm{tr}}$, assigning a probability to each scenario.

In practice, the stochastic process is inferred from historical data, represented by a matrix $\mathbf{Y}$ of size $d \times p$, where $d$ denotes the number of observed realizations of the stochastic process and $p$ is the dimension of each realization.

To the best of our knowledge, \pkg{scentree} is the first \proglang{Python} framework specifically designed for scenario fan and tree generation using statistical and machine learning algorithms. The framework is also well suited to two-stage stochastic programming problems, as it provides a systematic approach for generating data-driven representations of uncertainty from historical observations.

The remainder of this paper is organized as follows. Section \ref{sec:literature_review} presents and discusses the state of the art in scenario generation software. Section \ref{sec:scentree} provides a detailed description of the \pkg{scentree} package, followed by an illustrative example in Section \ref{sec:illustrative_example}. Finally, Section \ref{sec:conclusions_future_research} presents the conclusions and discusses potential directions for future research.

\section{Literature review}
\label{sec:literature_review}
As scenarios are a key component of multistage stochastic programming, several research works have focused on developing tools for scenario generation. Accordingly, the literature reviewed in this section is organized from two complementary perspectives. First, we discuss frameworks specifically developed for scenario generation. We then briefly review stochastic programming software that, while primarily focused on modeling and solving stochastic programs, incorporates tools to assist with scenario generation.

Recent advances in scientific computing have established \proglang{Julia} \parencite{Julia2017} as a promising programming language for data science, particularly for computationally intensive statistical and numerical methods. In this context, \textcite{Kirui2020} present the first \proglang{Julia} package dedicated to scenario generation for multistage stochastic programming, \pkg{ScenTrees.jl}. Scenario generation can be based either on a user-defined stochastic process or, when historical data are available, on kernel density estimation \parencite[KDE for short;][]{Silverman1986}, which is implemented in the package.

With regard to KDE, a limitation of the current implementation is that \code{kernel\_scenarios}, which is the functionality that generates trajectories, only supports univariate kernel density estimation. Its interface accepts only two-dimensional arrays (i.e. matrices), where rows represent sample paths and columns represent stages, so each observation is assumed to be scalar. As a result, dependencies among the components of multivariate random vectors cannot be modeled, preventing the direct generation of multivariate scenario trees from data.

\pkg{StochOptim} \parencite{Keutchayan2020} is a \proglang{Python} package for generating scenarios. It is a framework for solving two-stage and multistage stochastic problems. It includes algorithms for scenario tree construction. It supports both scenario generation from user-specified probability distributions and scenario clustering based on historical data. The official \textsf{GitHub} repository (\url{https://github.com/julienkeutchayan/StochOptim}) and the \textsf{PyPI} release (\url{https://pypi.org/project/stochoptim/}) have not been updated since 2021, suggesting that it is no longer actively maintained.

We next consider stochastic programming software primarily designed for modeling and solving stochastic programs that also provides some limited scenario generation capabilities. The work of \textcite{Knueven2023} presents \pkg{mpi-sppy}, a \proglang{Python} package whose main feature is its efficient parallelization capabilities when solving multistage stochastic optimization problems. This framework incorporates a functionality for defining and managing scenario trees, including scenario topology and  probabilities, non-anticipativity information, and parallel scenario processing, but does not provide dedicated methods for scenario generation.

\textcite{Biel2022} present \pkg{StochasticPrograms.jl}, an open-source framework for modeling and solving stochastic programming problems in \proglang{Julia}. The package provides an expressive modeling language, parallel implementations of structure-exploiting algorithms, such as variants of the L-shaped method, and distributed-memory support for large-scale instances. Regarding scenario generation, the framework does not provide methods for generating scenarios from historical data. Instead, it relies on user-supplied scenarios or on user-defined sampling mechanisms representing the underlying stochastic process.

\pkg{SDDP.jl} \parencite{Dowson2021} is an open-source \proglang{Julia} package for solving large multistage stochastic optimization problems using the Stochastic Dual Dynamic Programming (SDDP) algorithm \parencite{Philpott2013}. Built on top of \proglang{JuMP} \parencite{Dunning2017}, it provides a high-level modeling interface while maintaining performance comparable to low-level implementations. Its main contribution is the development of a flexible, solver-independent, and extensible implementation of SDDP. The package assumes that the uncertainty has already been discretized into a finite set of outcomes (or represented through a user-defined stochastic process). Consequently, \pkg{SDDP.jl} focuses on solving the resulting multistage stochastic program and does not provide methods for generating scenarios from historical data.

\section{Scentree package}
\label{sec:scentree}
The framework is inspired by \textcite{Heredia2018,Cuadrado2020,Sola2025}, who propose constructing scenario trees using time series factor analysis \parencite[TSFA;][]{Gilber2005}.

These works focus on modeling electricity price and variable renewable energy uncertainty within a multistage stochastic programming framework. The authors use TSFA to (i) obtain a low-dimensional representation of the data, (ii) estimate the underlying stochastic process, and (iii) predict future values to generate a scenario fan. Subsequently, tree reduction algorithms are applied to obtain a scenario tree, with the ultimate goal of supporting participation in the electricity market.

From an implementation perspective, \pkg{scentree} is designed to provide flexibility by allowing scenario fan generation to be performed independently of scenario tree construction. Thus, users can construct a scenario tree from a scenario fan generated with an external tool. The scenario fan generation methodology is presented in Section \ref{sec:fan_generation}, while the tree reduction procedure is described in Section \ref{sec:tree_reduction}. Section \ref{sec:output_structure} provides a detailed description of the package output.

Finally, note that the code is open-source, publicly available on \textsf{GitHub} at \url{https://github.com/mmobec/scentree}. It can be downloaded from \textsf{PyPI} at \url{https://pypi.org/project/scentree} and the documentation is on \textsf{readthedocs} at \url{https://scentree.readthedocs.io}.

\subsection{Scenario fan generation}
\label{sec:fan_generation}
As the dimension of the stochastic process, $p$, could be so large as to incur the curse of dimensionality \parencite{Bellman1957}, the first step is to obtain a low-dimensional representation by means of Principal Component Analysis \parencite[PCA;][]{Pearson1901}. The number of principal components is automatically selected to explain  at least 80\% of the total variance, following a commonly used variance-retention criterion \parencite{Dunteman1989}. After applying PCA to the data, the resulting matrix is of size $d \times k$, where $k \ll p$ denotes the dimension of the resulting space. Note that before applying PCA, the data are scaled so that the mean and variance of each column are 0 and 1, respectively. Let $\tilde{\mathbf{Y}}$ be the resulting matrix after applying PCA.

Once the data are preprocessed, a set of predictive model families, denoted by $\Gamma$, is considered to identify the best-performing model. The families in $\Gamma$ include multivariate time series models, such as vector autoregressive models \parencite{Sims1980}, and machine learning models, including support vector machines \parencite{Cortes1995}.

In order to select the best predictive model for each family in $\Gamma$, the first step consists of hyperparameter tuning. Since the data are indexed by $\mathcal{D}$, we use time series cross-validation \parencite{Bergmeir2012}. As a result, a set of predictive models, $\tilde{\Gamma}$, is obtained. Each model in $\tilde{\Gamma}$ is the best-performing one within its respective family.

Afterwards, the final predictive algorithm, $\gamma$, is selected using the root mean squared error:
\[
\gamma \coloneqq \argmin_{g \in \tilde{\Gamma}} \sqrt{\frac{1}{(d-h)k} \sum_{i=1}^{d-h} \Big\lVert g\!\left(\tilde{\mathbf{y}}_i, \tilde{\mathbf{y}}_{i+1}, \ldots, \tilde{\mathbf{y}}_{i+h-1}\right) - \tilde{\mathbf{y}}_{i+h} \Big\rVert_2^2}
\]
where $h$ denotes the number of lags, which is a hyperparameter selected in the first step, $\lVert \cdot\rVert_2$ the $\ell^2$ norm and $\tilde{\mathbf{y}}_j$ the $j-$th row of $\tilde{\mathbf{Y}}$.

Now, let $n_f$ denote the number of scenario fans (each associated with a distinct instance of the problem formulated in Equation \ref{eq:general_msop}) and $n_\alpha$ the number of scenarios in each fan required by the user. We derive $n_f$ predicted values using the predictive model $\gamma$. Note that the user may be interested in obtaining in-sample predictions, that is, estimating the last $n_f$ steps of the data. The package has a parameter, \code{build\_in\_sample\_fans}, that allows the user to choose whether the estimated values correspond to the last $n_f$ steps, i.e. from $d - n_f + 1$ to $d$, or to the next $n_f$ steps, i.e. from $d+1$ to $d + n_f$. Let $\{ \hat{\tilde{\mathbf{y}}}_1, \ldots, \hat{\tilde{\mathbf{y}}}_{n_f}\!\}$ be the set of predicted values.

Scenario fans are constructed by combining predicted values with estimated residuals. The latter are computed from historical data as
\[
    \hat{\mathbf{r}}_j \coloneqq \tilde{\mathbf{y}}_j - \hat{\tilde{\mathbf{y}}}_j
\]
where
\[
    \hat{\tilde{\mathbf{y}}}_j \coloneqq \gamma\!\left(\tilde{\mathbf{y}}_{j-h}, \ldots, \tilde{\mathbf{y}}_{j-1}\right)
\]
denotes the predicted value for $h+1 \leq j \leq d$. To obtain $n_f$ scenario fans, we follow the iterative process described in Algorithm \ref{alg:scenario_fan}. Note that the output is a tensor $\tilde{\mathbf{F}}$ of size $n_f \times n_\alpha \times k$ or, equivalently, a vector of length $n_f$ whose entries are matrices of size $n_\alpha \times k$. Essentially, to obtain a scenario fan, we apply a Monte Carlo method \parencite{Hammersley1964}.

\begin{algorithm}
\caption{Scenario fan generation}
\label{alg:scenario_fan}
\begin{algorithmic}
    \Require Set of predicted values $\{\hat{\tilde{\mathbf{y}}}_1,\ldots,\hat{\tilde{\mathbf{y}}}_{n_f}\}$
    \Require Set of residuals $\{\hat{\mathbf{r}}_{h+1},\ldots,\hat{\mathbf{r}}_d\}$
    \Require Number of scenarios $n_\alpha$
    \State Define a tensor $\tilde{\mathbf{F}}$ of size $n_f \times n_\alpha \times k$ and initialize it as empty
    \For{$a = 1,\ldots,n_f$}
        \State Sample, at random and without replacement, $n_\alpha$ residuals, $\{\hat{\mathbf{r}}_{\pi_1}, \ldots, \hat{\mathbf{r}}_{\pi_{n_\alpha}}\!\}$
        \State Define a matrix $\tilde{\mathbf{A}}$ of size $n_\alpha \times k$ and initialize it as empty
        \For{$b = 1,\ldots,n_\alpha$}
            \State $\tilde{\mathbf{A}}[b, \colon] \coloneqq \hat{\tilde{\mathbf{y}}}_a + \hat{\mathbf{r}}_{\pi_b}$
        \EndFor
        \State $\tilde{\mathbf{F}}[a, \colon, \colon] \coloneqq \tilde{\mathbf{A}}$
    \EndFor
    \State \textbf{Output:} Scenario fans in the low-dimensional space, i.e., tensor $\tilde{\mathbf{F}}$
\end{algorithmic}
\end{algorithm}

Due to the nature of PCA, any point in the low-dimensional space, whether or not it belongs to the historical data, can be mapped back to the original $p$-dimensional space. In brief, let $\mathbf{V}_k$ denote the matrix containing the $k$ principal directions selected during PCA. The original-space representation is then obtained by multiplying the low-dimensional representation by $\mathbf{V}_k^\top$. Accordingly, we define $\mathbf{F}$ as the tensor of size $n_f \times n_\alpha \times p$ containing the $p$-dimensional scenario fans obtained from $\tilde{\mathbf{F}}$ by (i) mapping them back to the original space and (ii) rescaling them to recover the initial data scale.

The first implementation includes a penalized linear regression model, a support vector machine, and a vector autoregressive model. We use \pkg{scikit-learn} \parencite{Pedregosa2011} and \pkg{statsmodels} \parencite{Seabold2010} to incorporate these predictive models. The framework also allows additional models from these libraries, as well as models from other libraries or custom implementations, to be integrated straightforwardly, as its design is agnostic to any particular package or implementation.

Finally, from a theoretical point of view, applying a dimensionality reduction transformation to a stochastic process results in another stochastic process, now taking values in a low-dimensional space. In particular, if $\beta \colon \mathbb{R}^p \to \mathbb{R}^k$ denotes the dimensionality reduction transformation (PCA in our case), the resulting process is defined as
\[
\tilde{\boldsymbol{\xi}}(\delta,\omega) \coloneqq \beta\big(\boldsymbol{\xi}(\delta,\omega)\big),
\qquad
\tilde{\boldsymbol{\xi}} \colon \mathcal{D} \times \Omega \to \mathbb{R}^k.
\]
Note that the index set $\mathcal{D}$ remains unchanged. This approach is also exploited by \textcite{Hyndman2009}, who forecast the principal component scores of functional observations as time series.

A scenario fan consists of complete trajectories but does not explicitly represent the information available at each stage. A scenario tree organizes these trajectories according to their shared history, providing a representation of the evolving uncertainty required in multistage stochastic programming \parencite{Shapiro2009}. In addition, the total number of nodes in the scenario tree is significantly smaller than in the scenario fan, reducing the size of the multistage stochastic programming problem to be solved while preserving representativeness of the underlying stochastic process. In Section \ref{sec:tree_reduction}, we describe how to construct the scenario tree $\boldsymbol{\xi}_{\mathrm{tr}}$ from the scenario fan.

\subsection{Scenario tree reduction}
\label{sec:tree_reduction}
\textcite{Heitsch2009} propose the forward tree construction (FTC) algorithm to construct a scenario tree from an initial scenario fan. The objective is to represent the scenario fan as a tree whose nodes represent possible realizations of the stochastic process and whose edges connect realizations across consecutive stages. Figure \ref{fig:tree_example} illustrates an example of a scenario tree for a three-stage multistage stochastic problem.

The procedure starts with a single root node to which all scenarios are associated. The tree is then constructed sequentially over the $s$ stages. Given a node at stage $1 \leq i < s$, the scenarios associated with that node are used to construct its successor nodes at stage $i+1$. This is done by selecting representative scenarios according to a prescribed distance criterion and subsequently redistributing the scenarios among the selected representatives.

The forward selection procedure iteratively selects representative scenarios from the set of scenarios associated with a node. Once the representatives have been selected, an optimal redistribution assigns each scenario to the representative that minimizes the corresponding distance. Each representative and its associated scenarios then define a successor node at the next stage. This procedure is applied recursively to the resulting nodes until the final stage is reached, yielding the scenario tree $\boldsymbol{\xi}_{\mathrm{tr}}$.

\begin{figure}
    \centering
    \begin{tikzpicture}[
        circlenode/.style={draw,circle,inner sep=0pt,minimum size=10pt,align=center},
        probnode/.style={right=20pt},
        every child node/.style={circlenode},
        edge from parent/.style={draw,->},
        level distance=50pt,
        level 3/.style={sibling distance=15pt}
    ]
        \node[circlenode] (root) {}[grow'=right]
            child [sibling distance=45pt] { node (s1) {}
                child { node (s2) {}
                    child { node (s3) {} }
                    child { node {} }
                }
            }
            child { node {}
                child [sibling distance=30pt] { node {}
                    child { node {} }
                    child { node {} }
                }
                child [sibling distance=30pt] { node {}
                    child { node {} }
                    child { node {} }
                }
            }
            child [sibling distance=37.5pt] { node {}
                child [sibling distance=0pt] { node {}
                    child { node {} }
                }
                child [sibling distance=30pt] { node {}
                    child { node {} }
                }
            };
        \def\ycord{-70pt}
        \path[font=\sffamily\small]
            let \p0 = (root) in (\x0,\ycord) node{Root}
            let \p1 = (s1) in (\x1,\ycord) node{Stage 1}
            let \p2 = (s2) in (\x2,\ycord) node{Stage 2}
            let \p3 = (s3) in (\x3,\ycord) node{Stage 3};
    \end{tikzpicture}
    \caption{Example of a scenario tree.}
    \label{fig:tree_example}
\end{figure}

It is important to note that each node results from clustering a set of scenarios and therefore contains all the scenarios assigned to the corresponding cluster. In terms of data, the resulting scenario tree can be represented by a matrix of size \(n_\alpha \times p\). At each stage, scenarios assigned to the same node share the same data values for the columns corresponding to that stage. These values are given by the scenario selected as the representative of that node.

The FTC algorithm also computes the probability of each node at each stage throughout the scenarios. Initially, the scenarios are assigned their fan probability (by default, a uniform distribution over the set of scenarios is assumed). Once the nodes are obtained, the probability of each node is computed by adding the probabilities of the scenarios assigned to it. At the implementation level, since each node contains a unique scenario representative, the node's probability is assigned to its representative scenario, while the other scenarios belonging to the same node are assigned a probability of zero.

Our framework implements the FTC algorithm to construct the scenario trees. For each of the $n_f$ scenario fans, a corresponding scenario tree is obtained. To run the algorithm, the tensor $\mathbf{F}$ obtained in Section \ref{sec:fan_generation} is considered. In terms of data, after constructing all the trees, a tensor $\mathbf{T}$ of size $n_f \times n_\alpha \times p$ is obtained.

Although FTC is one of the earliest algorithms proposed, considerable research has been devoted to this topic. Examples include \textcite{Hewitt2022,Medina2020,Li2019}. Consequently, the tree reduction component in \pkg{scentree} is agnostic to the underlying algorithm, allowing alternative tree reduction methods to be implemented without affecting the existing ones. The choice of method is therefore left to the user.

\subsection{Output structure}
\label{sec:output_structure}
The construction of the scenario fan and the scenario tree generates several relevant data. In particular, two parts of the process produce data that require further explanation: the generation of the scenario fan and the construction of the scenario tree.

Regarding scenario fan generation, the output is a dictionary with the following structure:
\begin{itemize}
    \item \code{observed\_values}: when \code{build\_in\_sample\_fans} is set to \code{True}, this is an array containing $n_f$ arrays, each storing the observed values. Otherwise, this variable contains no data.
    \item \code{predicted\_values}:  an array containing $n_f$ arrays, each storing the values predicted by the predictive model.
    \item \code{scenarios}: an array containing $n_f$ matrices, each representing the $n_\alpha$ trajectories of the stochastic process.
\end{itemize}

A multistage stochastic programming problem can be formulated in two different ways. The first, which we refer to as the \emph{nodal formulation}, associates decision variables with the nodes of the scenario tree, thereby enforcing nonanticipativity implicitly. The second, which we refer to as the \emph{nonanticipativity formulation}, explicitly imposes nonanticipativity constraints to ensure the consistency of decisions across scenarios sharing the same history. 

To support both formulations, the tree output provides the information required to construct either representation of the optimization problem. Accordingly, the output consists of a \proglang{Python} dictionary with two keys: \code{nodal\_formulation} and \code{nac\_formulation}, corresponding to the nodal and nonanticipativity formulations, respectively.

The output of the nodal formulation consists of an array of length $n_f$, with each element containing the following information:
\begin{itemize}
    \item \code{nodes}: the nodes belonging to each stage, represented by a dictionary whose keys are stage identifiers and whose values are lists of node identifiers.
    \item \code{node\_scenarios}: the scenarios associated with each node, represented by a dictionary whose keys are node identifiers and whose values are lists of scenario identifiers.
    \item \code{parent}: the parent node associated with each node, represented by a dictionary whose keys are node identifiers and whose values identify their respective parent nodes.
    \item \code{probability}: the probability associated with each node, represented by a dictionary whose keys are node identifiers and whose values specify the corresponding probabilities.
    \item \code{random\_variable\_value}: the scenario data values associated with each node, represented by a dictionary whose keys are node identifiers and whose values are \pkg{NumPy} arrays \parencite{Harris2020}  containing the corresponding values from matrix $\mathbf{T}$ (see Section \ref{sec:tree_reduction}).
    \item \code{stage}: the stage associated with each node, represented by a dictionary whose keys are node identifiers and whose values indicate the corresponding stage.
\end{itemize}
Note that the example provided in Section \ref{sec:illustrative_example} uses the nodal formulation. By directly supporting this formulation, the scenario tree can be easily and seamlessly integrated into optimization modeling frameworks, as demonstrated throughout the example.

On the other hand, the output of the nonanticipativity formulation consists of an array of length $n_f$, where each element is a dictionary following the same structure:
\begin{itemize}
    \item \code{scenario\_probabilities}: an array of length $n_\alpha$ containing the final probability of each scenario. All scenarios belonging to the same node have a probability of zero, except for the representative scenario, whose probability is equal to that of the node it represents.
    \item \code{scenario\_tree\_data}: a matrix of size $n_\alpha \times p$ containing the scenario tree data. As explained in Section \ref{sec:tree_reduction}, all scenarios belonging to the same node share exactly the same data values, thereby ensuring nonanticipativity.
    \item \code{tree}: an array representing the tree, in which each element corresponds to a node and consists of a dictionary organized as follows:
    \begin{itemize}
        \item \code{key}: a unique identifier for the node. It is a \proglang{Python} tuple whose first element indicates the stage and whose second element identifies the node within that stage.
        \item \code{scenario\_ids}: an array containing the identifiers of the scenarios belonging to this node.
        \item \code{parent\_key}: a tuple containing the identifier of the parent node.
        \item \code{description}: a human-readable string providing a textual representation of the \code{key} field.
    \end{itemize}
\end{itemize}

\section{Illustrative example: the Flowergirl problem}
\label{sec:illustrative_example}
We consider the Flowergirl inventory problem of
\textcite[][Example 1.4]{Pflug2014}: a retailer orders before uncertain demand,
carries surviving stock forward, and covers shortages through more expensive
emergency purchases. The multistage decision process is
\[
    x_0 \longrightarrow \xi_1 \longrightarrow x_1 \longrightarrow \xi_2 \longrightarrow x_2 \longrightarrow \xi_3.
\]
where $x_i$ are the orders and $\xi_i$ the demand, The tree has a deterministic root and three observation stages, ending in settlement rather than another regular order. At every stage the selling price is 1, the purchase cost is $b=0.85$, the emergency cost is $u=1.25$ and the survival fraction is $\ell=0.5$.

Let {$\boldsymbol{\xi}_{\mathrm{tr}}$} be the scenario tree, with root node $n=0$, where at every node $n>1$ the random variable $\xi(n)$ is observed with probability $\pi_n$. $\mathcal N_i$ is the set of nodes $n$ at stage $i$, $p(n)\in\mathcal N_{i-1}$ is the parent of node $n\in\mathcal N_i$, and $s(n)$ is the stage of node $n$ (thus $s(n)=i$). $\mathcal O=\bigcup_{i=0}^{2}\mathcal N_i$ denotes the nodes from the root to the stage $i=2$ included, and $\mathcal B=\bigcup_{i=1}^{3}\mathcal N_i$ the nodes from stage $i=1$ to final stage, included.

At every node, after observing the demand $\xi(n)$, a regular order $x(n)\ge 0$ will be placed, generating a stock $\zeta(n)=\zeta(n)^+-\zeta(n)^-$, $\zeta_0=0$. $\zeta(n)^+$ is the inventory available at the next stage, while $\zeta(n)^-\ge 0$ is the shortage that must be supplied at the emergency cost $u$. As in \textcite[][]{Pflug2014} we assume that the final stock $\zeta(n)^+$, $n\in\mathcal N_3$, if some, is sold at at price $\ell$.  The corresponding MSP is
\begin{align*}
    \max\quad &\sum_{n\in\mathcal B}\pi_n\,\xi(n)
    -\sum_{n\in\mathcal O}\pi_n\,b\,x(n)
    -\sum_{n\in\mathcal B}\pi_n\,u\,\zeta(n)^-
    +\ell \sum_{n\in\mathcal N_3}\pi_n\zeta(n)^+,
   \\[2pt]
    \text{s.t.}\quad &x(p(n))+\ell\,\zeta(p(n))^+-\xi(n)
    =\zeta(n)^+-\zeta(n)^-\ \qquad n\in\mathcal B,
    \\[-2pt]
    &\zeta(n)^+,\zeta(n)^-\ge0 \ \,\qquad \qquad \qquad \qquad \qquad \qquad \qquad n\in\mathcal B,\\
    &x(n)\ge0\qquad \qquad \qquad  \qquad \qquad \qquad \qquad \qquad \qquad n\in\mathcal O,
   \\[-2pt]
    &\zeta(0)^+=0. 
\end{align*}

Throughout this section, we provide a set of code listings illustrating the different steps of the proposed workflow and enabling readers to reproduce the results presented. These listings cover the main functionalities of the \pkg{scentree} package, from scenario fan generation to scenario tree construction and its subsequent use in the optimization problem. For completeness, the complete implementation, including all the code required to reproduce the results, is available at \url{https://github.com/mmobec/scentree/blob/main/scentree/examples/flowergirl.ipynb}.

\subsection{Data loader}
Our package includes demand data for 401 consecutive days, represented by a $401 \times 3$ matrix. Each row represents a three-period campaign, with consecutive rows corresponding to campaigns on consecutive days rather than demand periods. We solve the optimization problem for the final day, corresponding to row 401, which we refer to as the {\em target day}.

Accordingly, we partition the data into two sets: the historical data, consisting of the first 400 days and denoted by $\mathbf{Y}$ throughout this work, and the data corresponding to the target day. The historical data are used to generate the scenario fan for the target day (Section \ref{sec:pkg_fan_generation}), from which a reduced scenario tree is constructed (Section \ref{sec:pkg_tree_reduction}) and subsequently used to solve the multistage stochastic programming problem (Section \ref{sec:pkg_solution}). 

Listing \ref{lst:data_loader} illustrates how the data are imported, loaded and partitioned. The \code{DatasetLoader} class provides methods for organizing and manipulating data. It is particularly useful when working with multiple datasets, as it facilitates their management. As shown in Listing \ref{lst:data_loader} the data corresponding to the target day are stored in \code{target\_day} variable.

\begin{lstlisting}[
    caption={Data import and loading.},
    language=Python,
    numbers=none,
    label={lst:data_loader},
]
import numpy as np
from scentree.io.data import load_flowergirl
from scentree.io.loader import Dataset, DatasetsLoader

# Load the data and convert it into a scentree dataset
flowergirl_ds = load_flowergirl()
Y, target_day = flowergirl_ds[:-1],flowergirl_ds[-1]
datasets = [Dataset(
    name="demand",
    values=Y,
    stage_ids=[1, 2, 3],
    bounds=(0., np.inf),
)]
dsl = DatasetsLoader(datasets=datasets)
full_values = dsl.get_full_values()
stage_ids = dsl.get_sorted_stage_ids()
num_vars = dsl.get_num_variables_per_stage()
stage_cols_map = dsl.create_stage_columns_mapping()
\end{lstlisting}

Figure \ref{fig:flowergirl_dataset} shows the historical demand data for 400 consecutive days, with three demand periods per day. The black dashed line represents the demand observations for the target day. This representation highlights the variability in demand across both days and periods within each campaign.

\subsection{Fan generation}
\label{sec:pkg_fan_generation}
The scenario fan can be generated as illustrated in Listing \ref{lst:fan_generator}. \code{StageManager} class is responsible of fan generation. The user must specify the number of fans and scenarios; in our case, these are set to 1 and 80, respectively.
\begin{lstlisting}[
    caption={Fan generator.},
    language=Python,
    numbers=none,
    label={lst:fan_generator},
]
from scentree.fan_generator import StageManager

num_fans = 1
num_scenarios = 80
stage_manager = StageManager()
scenario_fans = stage_manager.generate_scenario_fans(
    X=full_values,
    num_fans=num_fans,
    num_scenarios=num_scenarios,
    build_in_sample_fans=False,
    seed=204
)
fan = scenario_fans["scenarios"][0]
\end{lstlisting}

Furthermore, the framework can generate scenario fans for both observed and unobserved days, i.e., future days. This behavior is controlled by the boolean variable \code{build\_in\_sample\_fans}. For observed days, the scenario fans correspond to the last \code{num\_fans} days, whereas for future days, they correspond to the subsequent \code{num\_fans} days. In both cases, the scenario fans are generated in increasing chronological order. In our case,  \code{build\_in\_sample\_fans} is set to false.

Figure \ref{fig:flowergirl_fan} shows the scenario fan generated for the target day, consisting of 80 demand trajectories. Each trajectory represents a possible realization of demand across the three periods of the campaign. The black dashed line represents the actual demand for the target day.

\subsection{Tree reduction}
\label{sec:pkg_tree_reduction}
We next employ the package to construct a reduced tree. Listing \ref{lst:scenario_tree} illustrates how to use the package for this purpose. The output is stored in \code{tree\_instance}, a list of dictionaries, each representing a scenario tree for a different day. In this example, the list contains a single dictionary because only one day is requested. The tree reduction algorithm reduces the 80-trajectory fan to nine terminal tree scenarios, producing a 1--4--7--9 tree.
\begin{lstlisting}[
    caption={Tree reduction.},
    language=Python,
    numbers=none,
    label={lst:scenario_tree},
]
from scentree.tree_construction.ftc import FTC

tree_instance = FTC(
    scenarios=scenario_fans["scenarios"],
    num_variables_per_stage=num_vars,
    stage_ids=stage_ids
)
tree_structure = tree_instance.generate_scenario_trees(
    r=1.,
    stage_columns_mapping=stage_cols_map,
    initial_stage_id_to_cluster=1,
    seed=15,
)
tree_nac = tree_structure["nac_formulation"][0]
tree_data = tree_nac["scenario_tree_data"]
tree_data_matrix = np.unique(tree_data, axis=0)
\end{lstlisting}

\begin{figure}
    \centering
    \begin{subfigure}{0.8\textwidth}
        \centering
        \includegraphics[width=\textwidth]{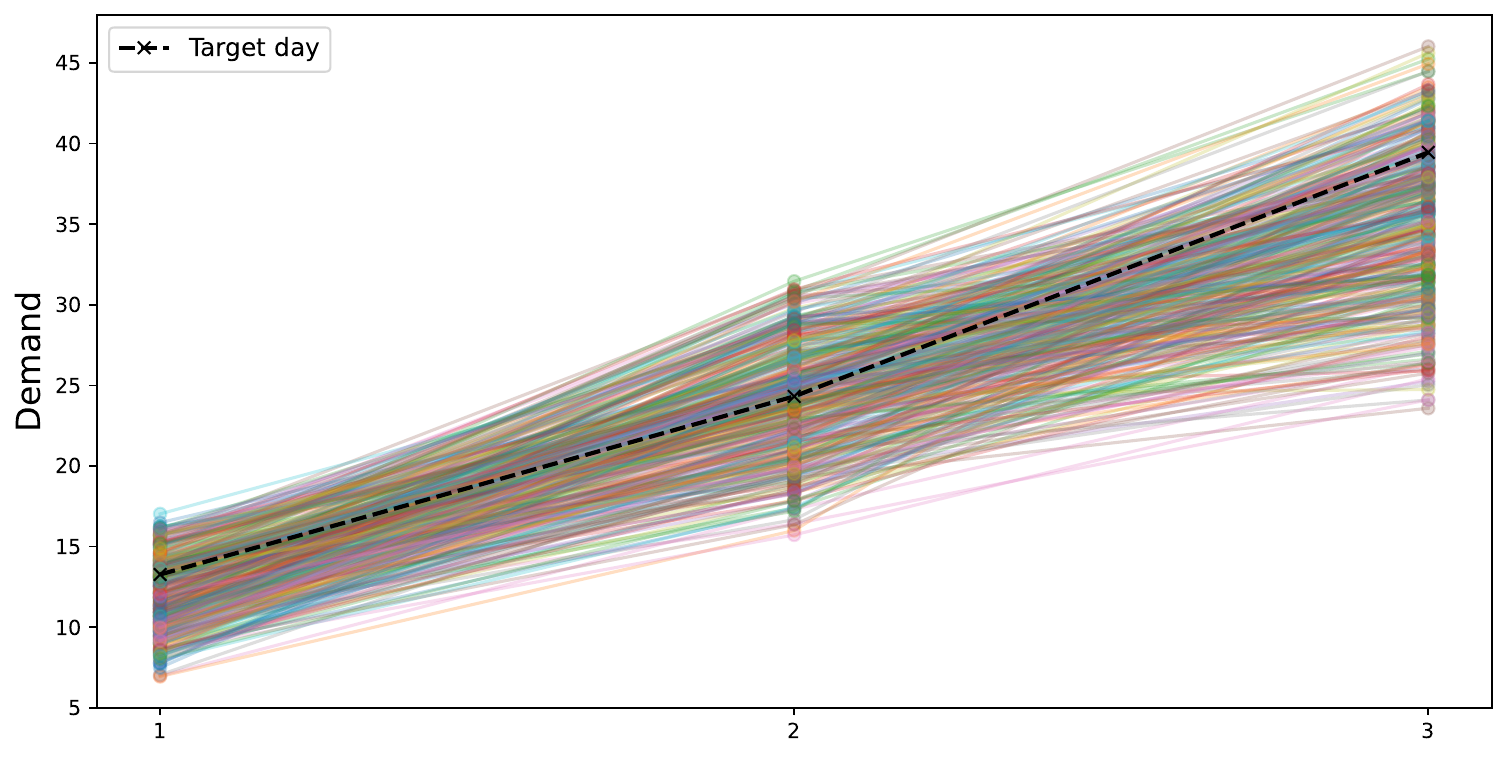}
        \caption{Historical data composed by 400 trajectories.}
        \label{fig:flowergirl_dataset}
    \end{subfigure}

    \begin{subfigure}{0.8\textwidth}
        \centering
        \includegraphics[width=\textwidth]{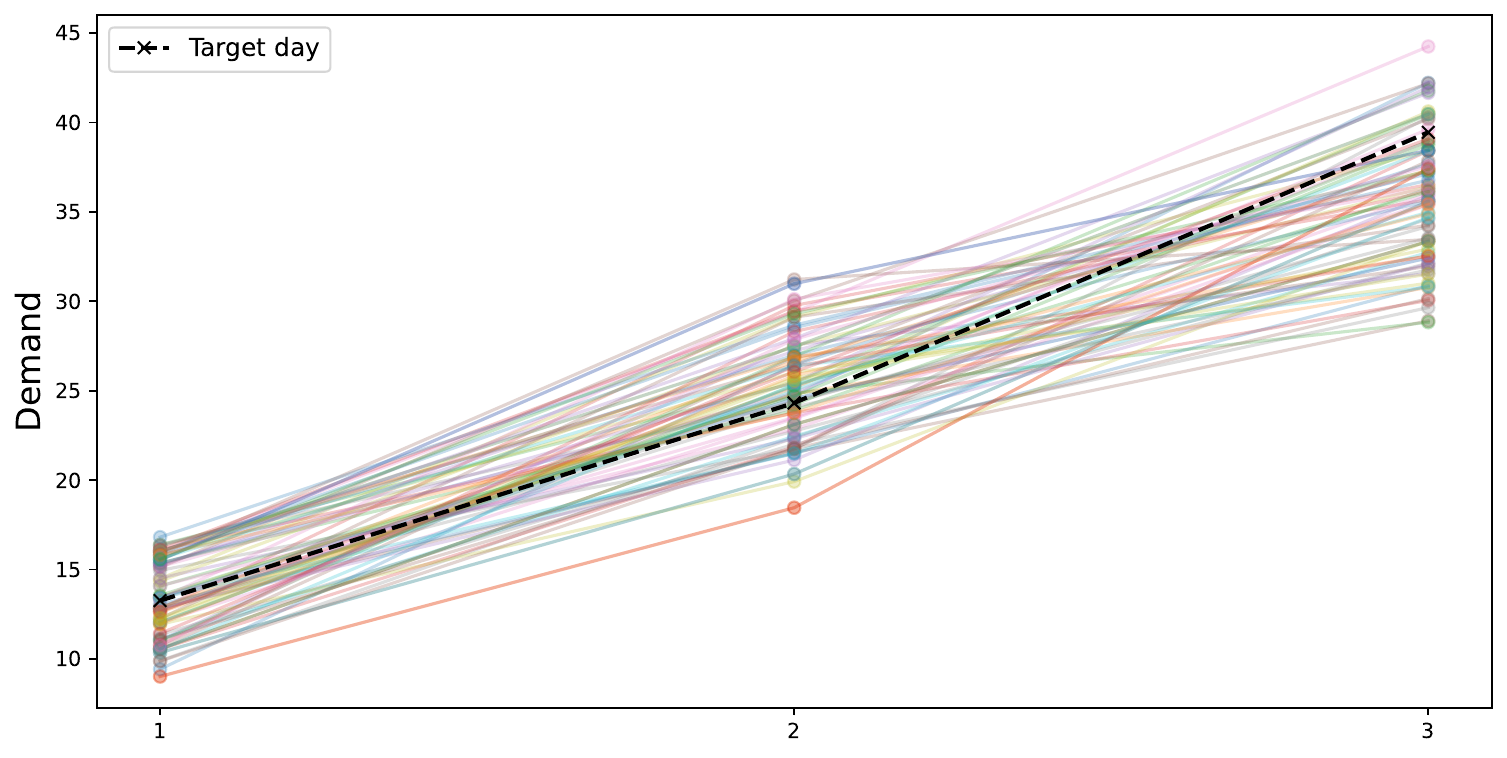}
        \caption{Generated scenario fan with 80 scenarios}
        \label{fig:flowergirl_fan}
    \end{subfigure}

    \begin{subfigure}{0.8\textwidth}
        \centering
        \includegraphics[width=\textwidth]{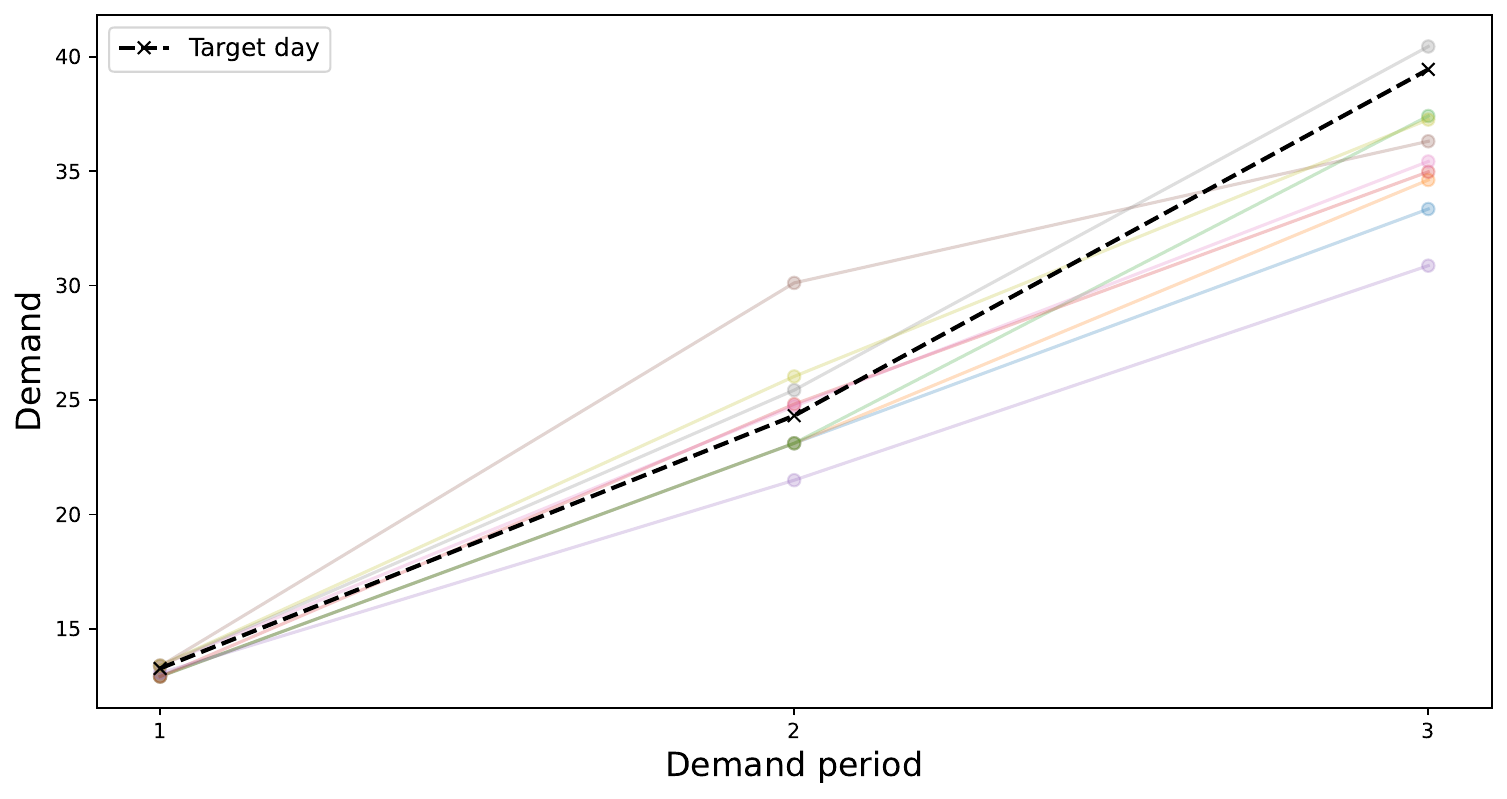}
        \caption{Reduced scenario tree containing nine scenarios.}
        \label{fig:flowergirl_tree}
    \end{subfigure}

    \caption{Historical demand data, scenario fan, and reduced scenario tree for the Flowergirl problem. The black dashed line indicates the target day for which the scenario fan and reduced scenario tree are constructed.}
    \label{fig:flowergirl}
\end{figure}

Figure \ref{fig:flowergirl_tree} shows the scenario tree obtained by reducing the scenario fan to nine scenarios for the target day. The black dashed line indicates the actual demand observations for the target day.

\subsection{Solution of the problem}
\label{sec:pkg_solution}
Finally, we illustrate how the resulting stochastic programming problem is solved. We use \pkg{Pyomo} \parencite{Hart2011} as the optimization modeling framework.  Before formulating the problem, we extract the data required from the reduced scenario tree, as shown in Listing \ref{lst:preprocessing_pyomo}.

\begin{lstlisting}[
    caption={Data preprocessing for Pyomo compatibility defining the data structure of the Flowergirl formulation.
    },
    language=Python,
    numbers=none,
    label={lst:preprocessing_pyomo},
]
ndl_formulation = tree_structure['nodal_formulation'][0]
parent = ndl_formulation["parent"]
stage = ndl_formulation["stage"]
xi = ndl_formulation["random_variable_value"]
pi = ndl_formulation["probability"]
members = ndl_formulation["node_scenarios"]
nodes = ndl_formulation['nodes']
\end{lstlisting}

\begin{figure}
    \centering
    \includegraphics[width=0.96\textwidth]{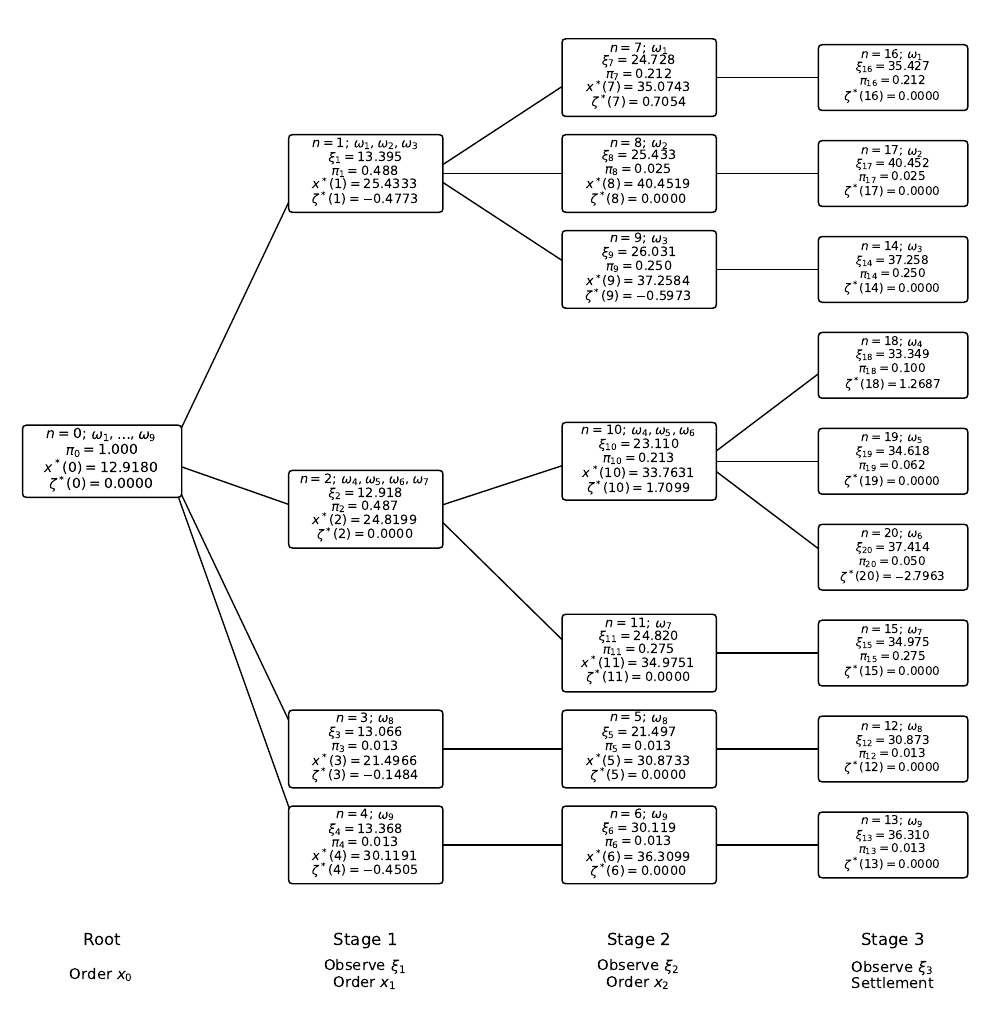}
    \caption{Flowergirl tree represented as a directed graph. Every node includes: node number $n$, demand $\xi_n=\xi(n)$, $\pi_n$, $x^*(n)$ and $\zeta^*(n)$ and the scenarios $\omega_i$ traversing the node.}
    \label{fig:flowergirl_graph}
\end{figure}

The model incorporates the reduced scenario tree and its associated probabilities to account for demand uncertainty across the three periods. The implementation is provided in Listing \ref{lst:pyomo_values}. The values of the  optimal initial order and the optimal expected profit are $x_0^*=12.9179$ and $10.5686$, respectively. Figure \ref{fig:flowergirl_graph} illustrates the scenario tree in graph form and the optimal solution $x^*$ and $\zeta^*$.

This completes the workflow, from scenario fan generation and scenario tree reduction using \pkg{scentree} to the formulation and solution of the optimization problem using \pkg{Pyomo}. By integrating these steps into a single process, the package provides a systematic approach to incorporating demand uncertainty into the optimization model.
\begin{lstlisting}[
    caption={Pyomo implementation of the Flowergirl optimization problem.},
    language=Python,
    numbers=none,
    label={lst:pyomo_values},
    showstringspaces=false,
]
import pyomo.environ as pyo
def flowergirl_model(parent, stage, xi, pi):
    root = next(n for n in stage if parent[n] is None)
    N = list(stage)
    B = [n for n in N if n != root]
    O = [n for n in N if stage[n] < 3]
    F = [n for n in N if stage[n] == 3]
    m = pyo.ConcreteModel()
    m.b, m.u, m.ell = 0.85, 1.25, 0.5
    m.x = pyo.Var(O, domain=pyo.NonNegativeReals)
    m.zeta_plus = pyo.Var(
        N, domain=pyo.NonNegativeReals
    )
    m.zeta_minus = pyo.Var(
        B, domain=pyo.NonNegativeReals
    )
    m.zeta_plus[root].fix(0.)
    m.zeta = pyo.Expression(
        B,
        rule=lambda m, n:m.zeta_plus[n] - m.zeta_minus[n]
    )
    def balance(m, n):
        a = parent[n]
        prod = m.x[a] + m.ell * m.zeta_plus[a]
        return prod - xi[n] == m.zeta[n]
    m.balance = pyo.Constraint(B, rule=balance)
    m.profit = pyo.Objective(
        sense=pyo.maximize,
        expr=sum(pi[n] * xi[n] for n in B)
        - m.b * sum(pi[n] * m.x[n] for n in O)
        - m.u * sum(pi[n] * m.zeta_minus[n] for n in B)
        + m.ell * sum(pi[n] * m.zeta_plus[n] for n in F)
    )
    return m

model = flowergirl_model(parent, stage, xi, pi)
solver = pyo.SolverFactory("appsi_highs")
solution = solver.solve(model)
\end{lstlisting}

\section{Conclusions and future research}
\label{sec:conclusions_future_research}
This work presents \pkg{scentree}, an open-source \proglang{Python} package for generating scenario fans and scenario trees for multistage stochastic programming using historical data. To the best of our knowledge, \pkg{scentree} is the first \proglang{Python} package specifically designed to generate scenario fans and scenario trees using machine learning and multivariate time series models for multivariate stochastic processes. The framework automates the main steps of generation process, from dimensionality reduction and model selection to scenario fan generation and scenario tree construction.

\pkg{Scentree} has three main features that highlight the value of its contribution. First, it does not require the user to specify a probability distribution for the underlying stochastic process, and the selection of predictive models is performed automatically. Consequently, only limited knowledge of statistics and machine learning is required to generate scenarios from historical data. Second, the framework is flexible by design. It is agnostic to the particular predictive model or tree construction method, allowing new models and scenario tree construction or clustering methods to be incorporated without affecting the existing implementations. In addition, its modular design allows partial executions of the framework. Third, the joint modeling of the multivariate stochastic process naturally captures the correlation between random variables.

Several extensions are planned for future research. First, we intend to incorporate additional machine learning, with particular attention to diffusion models and other recent techniques for time series modeling. These models may provide alternative approaches for capturing complex temporal dependencies and generating scenario fans. Second, recent methodologies for scenario tree construction will be investigated and incorporated into the framework. Finally, we plan to establish a connection with functional data analysis, which could provide a natural framework for modeling high-dimensional stochastic processes and their temporal evolution.

Overall, \pkg{scentree} is intended to provide a flexible and accessible framework for scenario generation, while maintaining sufficient extensibility to accommodate future developments in predictive modeling and scenario tree construction.

\section*{Acknowledgments}
This work has been supported with grants PID2022-139219OB-I00 and Cetp-FP-2023-00185 from the Spanish Ministerio de Ciencia, Innovación y Universidades. This research has been funded by CETPartnership, the Clean Energy Transition Partnership under the 2023 joint call for research proposals, co-funded by the European Commission (GA N°101069750) and with the funding organisations FFG (Austria), AEI (Spain) and MUR (Italy).

\printbibliography
\end{document}